\documentclass[a4paper, 11pt, one column, aas_macros]{article}

\usepackage[english]{babel}
\usepackage[utf8x]{inputenc}
\usepackage[T1]{fontenc}
\usepackage{aas_macros}
\usepackage{multicol}
\usepackage{booktabs} 
\usepackage{array}    
\usepackage[top=0cm, bottom=2.0cm, outer=2.5cm, inner=2.5cm, heightrounded,
marginparwidth=1cm, marginparsep=1cm, margin=1.5cm]{geometry}

\usepackage{caption}
\usepackage{graphicx} 
\graphicspath{ {./Images/} }
\usepackage{hyperref}
\usepackage{amsmath}  
\usepackage{amsfonts} %
\usepackage{amssymb}  %
\usepackage{multirow}
\usepackage{caption}
\usepackage{subcaption}
\usepackage{graphicx}
\usepackage{csquotes}

\title{Increasing Efficiency of the Chain of Contagion Task}
\author{{\rm Lars OM Rothkegel}\\Universität Potsdam\\
\and {\rm Jakob Fink-Lamotte}\\Universität Potsdam}

\date{}
\begin{document}
\maketitle
\begin{abstract}
The chain of contagion task (CCT) is a pychological test to measure the amount of contagious beliefs in individuals. Contagious beliefs thereby refer to the perception that certain objects, people, or substances can transmit contamination through mere contact or proximity  \cite{rozin1986operation}. In the CCT, a neutral object (usually a pen) is rubbed against an inherently disgusting object (e.g. a toilet paper with feces) and participants are asked how contaminated this pen is on a scale from 0 (not at all) to 100 (very contaminated). Afterwards, this pen is rubbed against another pen, and again, the experienced degree of contamination is assessed. This is repeated 12 times. The CCT has first been experimentally investigated by Tolin et al. (2004) in an in vivo procedure with real disgusting objects. The authors could show that contagious beliefs measured with the CCT show a strong bias for people with contamination-based obsessive-compulsive disorder (C-OCD) compared to anxious individuals and non-anxious controls. Fink-Lamotte et al. (2024) replicated these findings with an online version of the CCT using audio-imagery-based and video-based stimuli and instructions. Both studies used 12 pens to assess the degree of contagious beliefs. Within this brief report, we show that after 8 pens, hardly any additional variance is explained between participants and after the tenth pen, no new information is gained. Thus, we recommend only using 8 pens instead of 12 when using the CCT to assess contagious beliefs.

\end{abstract}

\noindent

\section{Introduction}
Disgust is a basic emotion that motivates avoidance of potentially contaminating stimuli and is characterized by a strong sense of revulsion, often accompanied by nausea \cite{rozin1987perspective}. In contamination-based obsessive-compulsive disorder (C-OCD), heightened disgust sensitivity contributes to exaggerated contagious beliefs and compulsive behaviors aimed at neutralizing perceived threats \cite{eyal2021disgust}. Contagious beliefs, as conceptualized by Rozin et al. (1986), reflect the notion that contact with a disgusting object can result in lasting contamination, even in the absence of physical threat. Understanding and evaluating contagious beliefs is therefore crucial for improving treatment approaches and deepening our insight into the mechanisms that maintain C-OCD. The Chain of Contagion Task (CCT), originally introduced by Tolin et al. (2004), provides a behavioral measure of contagious beliefs by assessing how contamination is perceived to spread from an initial object to a sequence of clean items. While the original paradigm relied on physical disgust-related objects and was limited in practical application, Fink-Lamotte et al. (2024) recently developed and validated an imagery- and video-based version of the CCT suitable for both experimental and clinical contexts. Their findings replicated that the imagery-based CCT reliably distinguishes individuals with C-OCD from both anxious and non-anxious controls, based on their persistent contamination ratings across a 12-pen sequence. However, the task's length may pose a burden for clinical and research use. To address this, we sought to determine the minimum number of pens required to retain the task’s discriminatory power.

\section{Method} \label{SecMethod}
To assess the variance explained by each pen we used the data from the study by Fink-Lamotte et al. (2024) 
with N=168 participants who performed the imagery- or video-based version of the CCT. Please note that this sample is larger than the sample in the original article, because additional data was collected after analysis and preparation of the manuscript. All information about the study procedure can be found within the \href{https://www.sciencedirect.com/science/article/pii/S0005789424000376?casa_token=IYtaAAAuQQ0AAAAA:TEG0bcTbRQu9uEaY6IpY_UsU5TQqmUyVJdT1sJOT2HTJFMc2_x_5w1fePlQ1-8XGnvzd-Z5YTA}{original article}. For the test of efficiency, the goal was to find the first pen in the chain, which does not explain any additional information regarding differences in contagious beliefs between participants. To identify this pen, linear mixed models were conducted with contamination ratings as the dependent variable, pen as a fixed effect, and subject as a random effect. Contrasts were specified such that each pen was compared to the subsequent one in the order of presentation in the CCT. If a pen showed no predictive power compared to the previous pen, new contrasts were defined by comparing the first non-significant pen to all subsequent pens. The first pen in this iterative process for which all subsequent comparisons are non-significant marks the endpoint of the procedure. As a second indicator of the amount of new information each pen provides, we computed Spearman correlations among all 12 pens to assess the consistency of rank order across pens. All analysis were conducted in R \cite{R}, with the addition of packages "lme4" \cite{bates2015package}, "lmerTest" \cite{kuznetsova2015package}, "ggplot2" \cite{wickham2016package} and "multcomp" \cite{hothorn2016package}. The according scripts and data can be found in an \href{https://osf.io/np73c/?view_only=831f65a674a34fdc98599d05780da6e4}{OSF repository}.

\section{Results}
The first linear mixed model showed that pens 2--5 each produced significantly different contamination ratings compared to their respective predecessors. Thus, the first five pens in the CCT each contributed significantly to predicting contamination ratings. Differences for successive comparisons for pens 1--5 constantly decreased from comparing pen 1 to pen 2 ($t=-12.914$, $p<0.001$) to comparing pen 4 to pen 5 ($t=-5.40$, $p=0.004$) until the first non-significant comparison (pen 5 vs. pen 6, $t=-3.555$, $p=0.066$). Thus, we started the iterative comparison with pen 5. Comparing pen 5 to all other pens after pen 6 led to significant differences (all $p_s<0.001$). Comparing pen 6 to all later pens was significant for pens 8 through 12 (pen 6 vs. pens 7-8: $p_s <0.01$, pen 6 vs. pens 9-12 $p_s<0.001$). Comparing pen 7 to the following was significant for pens 10 through 12 (pen 7 vs. pens 10-12 $p_s <0.05$, pen 7 vs. pen 12 $p_s<0.001$). Comparing pen 8 to the following was only significant when compared to pen 12 ($p=0.02$). Comparisons between pen 9 and each of the subsequent pens yielded no significant differences. Thus, pens 10, 11 and 12 provided no incremental explanatory power regarding contamination rating compared to pens 1--9. 

\begin{figure}
\includegraphics[width=1\textwidth]{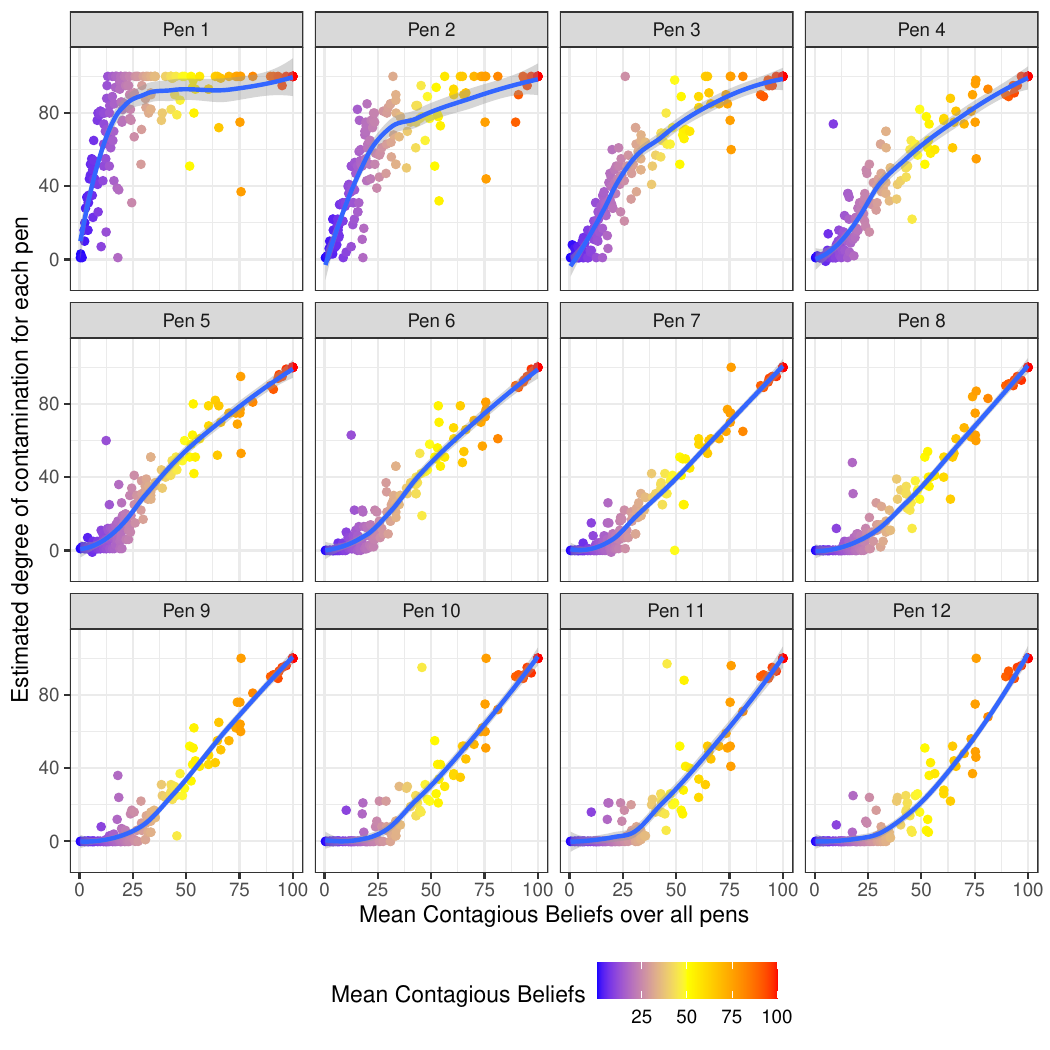}
\captionof{figure}{y-axis: Estimated contagious beliefs for each participant (N=168) and pen; x-axis: mean contagious beliefs over all 12 pens per participant sorted from 0 to 100. The subjects are ordered, such that values on the x-axis represent the mean contagious belief per participant over all 12 pens. On the y-Axis, the specific value for each pen is depicted. Thus, values on the left are on average low, values on the right are on average high and values in the middle on average have more movement between the pens.}
\label{CCT_Pen}
\end{figure}

To illustrate the distribution of assessed contamination for each pen, Figure \ref{CCT_Pen} shows the degree of contamination all 168 subjects attributed to each pen. This figure shows that all curves from pens 1--8 change their shape, but afterwards no apparent change is visible. This aligns with the linear mixed model results, which indicated that pens beyond pen 8 did not substantially contribute to differentiating between participants.

Spearman correlations (see Table \ref{Table_Cor}) ranged from $rho=0.43$ (pen 1 vs. pen 12) to $rho=0.98$ (pen 10 vs. pen 11). The cut-off pen was defined as the pen, which was correlated to the next one with $rho>0.95$, if this was true for all following pens. This cut-off identified pen 10 as the last pen providing relevant information. However, the correlation between pen 9 and 10 was $rho>0.949$ and the previous correlation between pen 8 and 9 was $rho>0.95$. Given the minimal deviation and the results from the linear mixed models, this provides additional evidence that pen 8 or 9 marks the point at which no further relevant information is gained. 

\begin{table}[h]
\centering
\caption{Spearman correlation Matrix between Pens}
\label{Table_Cor}
\begin{tabular}{l*{12}{>{\centering\arraybackslash}p{0.8cm}}}
\toprule
 & \textbf{Pen 1} & \textbf{Pen 2} & \textbf{Pen 3} & \textbf{Pen 4} & \textbf{Pen 5} & \textbf{Pen 6} & \textbf{Pen 7} & \textbf{Pen 8} & \textbf{Pen 9} & \textbf{Pen 10} & \textbf{Pen 11} & \textbf{Pen 12} \\
\midrule
\textbf{Pen 1}  & 1.00 & 0.82 & 0.73 & 0.64 & 0.59 & 0.54 & 0.54 & 0.54 & 0.47 & 0.47 & 0.45 & 0.43 \\
\textbf{Pen 2}  &       & 1.00 & 0.90 & 0.81 & 0.75 & 0.72 & 0.74 & 0.69 & 0.66 & 0.66 & 0.63 & 0.62 \\
\textbf{Pen 3}  &       &       & 1.00 & 0.93 & 0.88 & 0.86 & 0.85 & 0.82 & 0.79 & 0.78 & 0.76 & 0.74 \\
\textbf{Pen 4}  &       &       &       & 1.00 & 0.91 & 0.90 & 0.89 & 0.86 & 0.83 & 0.80 & 0.77 & 0.76 \\
\textbf{Pen 5}  &       &       &       &       & 1.00 & 0.97 & 0.93 & 0.93 & 0.91 & 0.86 & 0.84 & 0.82 \\
\textbf{Pen 6}  &       &       &       &       &       & 1.00 & 0.93 & 0.94 & 0.93 & 0.87 & 0.86 & 0.84 \\
\textbf{Pen 7}  &       &       &       &       &       &       & 1.00 & 0.94 & 0.91 & 0.90 & 0.88 & 0.86 \\
\textbf{Pen 8}  &       &       &       &       &       &       &       & 1.00 & 0.96 & 0.93 & 0.91 & 0.90 \\
\textbf{Pen 9}  &       &       &       &       &       &       &       &       & 1.00 & 0.95 & 0.94 & 0.93 \\
\textbf{Pen 10} &       &       &       &       &       &       &       &       &       & 1.00 & 0.98 & 0.96 \\
\textbf{Pen 11} &       &       &       &       &       &       &       &       &       &       & 1.00 & 0.97 \\
\textbf{Pen 12} &       &       &       &       &       &       &       &       &       &       &       & 1.00 \\
\bottomrule
\end{tabular}
\end{table}

\section{Discussion \& Conclusion}
With a post-hoc analysis of data from a rather differential sample performing the CCT, we aimed at increasing the tests efficiency. Analyses indicated that 8 (efficient), 9 (conservative) or 10 (very conservative) pens are enough to perform the task without losing discriminatory power. Information, which could be lost due to terminating the task after pen 8 concern participants who after the $7^{th}$ or $8^{th}$ pen either show an increase (which would presumably be result of incoherent or careless answering) or a sudden decrease of perceived contamination. A sharp decrease in contagious beliefs at such a late stage in the sequence appears implausible when inspecting responses across all pens and participants. Therefore, we are confident that using only the first 8 pens yields comparable results in distinguishing contagious beliefs between participants. 
\\ \\
The CCT is designed to assess the extent of contagious beliefs in participants, a disgust-related factor thought to play a key role in the maintenance of psychopathology such as C-OCD \cite{tolin2004sympathetic, rozin1986operation}. Improving its efficiency is particularly important if the test is used frequently, for instance by psychotherapists to identify potential treatment barriers in C-OCD. Contagious beliefs are a more cognitive aspects of OCD maintainance and very rigid and thus may not be targeted by the gold standard to treat OCD, exposure with response prevention \cite{hirschtritt2017obsessive}. Individually targeted psychotherapeutic treatments are key for optimizing outcome if ongoing treatmens fails (\cite{delgadillo2018feedback}). Thus, we believe this task can be a valuable addition to the standard diagnostic procedure for C-OCD and other disgust-related psychopathologies
\\ \\
To conclude, we reccomend using the \href{https://zenodo.org/records/7730459}{virtual version of the CCT} with 8 pens when aiming to assess the amount of contagious beliefs in individuals.

\bibliographystyle{plain}
\bibliography{main}

\begin{thebibliography}{10}

\bibitem{bates2015package}
Douglas Bates, Martin Maechler, Ben Bolker, Steven Walker, Rune Haubo~Bojesen Christensen, Henrik Singmann, Bin Dai, Gabor Grothendieck, Peter Green, and M~Ben Bolker.
\newblock Package ‘lme4’.
\newblock {\em Convergence}, 12(1):2, 2015.

\bibitem{delgadillo2018feedback}
Jaime Delgadillo, Kim de~Jong, Mike Lucock, Wolfgang Lutz, Julian Rubel, Simon Gilbody, Shehzad Ali, Elisa Aguirre, Mark Appleton, Jacqueline Nevin, et~al.
\newblock Feedback-informed treatment versus usual psychological treatment for depression and anxiety: a multisite, open-label, cluster randomised controlled trial.
\newblock {\em The Lancet Psychiatry}, 5(7):564--572, 2018.

\bibitem{eyal2021disgust}
Tal Eyal, Reuven Dar, and Nira Liberman.
\newblock Is disgust in obsessive-compulsive disorder mediated by fear of pathogens?
\newblock {\em Journal of Anxiety Disorders}, 77:102340, 2021.

\bibitem{hirschtritt2017obsessive}
Matthew~E Hirschtritt, Michael~H Bloch, and Carol~A Mathews.
\newblock Obsessive-compulsive disorder: advances in diagnosis and treatment.
\newblock {\em Jama}, 317(13):1358--1367, 2017.

\bibitem{hothorn2016package}
Torsten Hothorn, Frank Bretz, Peter Westfall, Richard~M Heiberger, Andre Schuetzenmeister, Susan Scheibe, and Maintainer~Torsten Hothorn.
\newblock Package ‘multcomp’.
\newblock {\em Simultaneous inference in general parametric models. Project for Statistical Computing, Vienna, Austria}, pages 1--36, 2016.

\bibitem{kuznetsova2015package}
Alexandra Kuznetsova, Per~Bruun Brockhoff, Rune Haubo~Bojesen Christensen, et~al.
\newblock Package ‘lmertest’.
\newblock {\em R package version}, 2(0):734, 2015.

\bibitem{R}
{R Core Team}.
\newblock {\em R: A Language and Environment for Statistical Computing}.
\newblock R Foundation for Statistical Computing, Vienna, Austria, 2023.

\bibitem{rozin1987perspective}
Paul Rozin and April~E Fallon.
\newblock A perspective on disgust.
\newblock {\em Psychological review}, 94(1):23, 1987.

\bibitem{rozin1986operation}
Paul Rozin, Linda Millman, and Carol Nemeroff.
\newblock Operation of the laws of sympathetic magic in disgust and other domains.
\newblock {\em Journal of personality and social psychology}, 50(4):703, 1986.

\bibitem{tolin2004sympathetic}
David~F Tolin, Patrick Worhunsky, and Nicholas Maltby.
\newblock Sympathetic magic in contamination-related ocd.
\newblock {\em Journal of behavior therapy and experimental psychiatry}, 35(2):193--205, 2004.

\bibitem{wickham2016package}
Hadley Wickham, Winston Chang, and Maintainer~Hadley Wickham.
\newblock Package ‘ggplot2’.
\newblock {\em Create elegant data visualisations using the grammar of graphics. Version}, 2(1):1--189, 2016.

\end{thebibliography}

\end{document}